# A Nanometer-Thick Oxide Semiconductor Transistor with Ultra-High Drain Current

Zehao Lin,[1] Mengwei Si,[1] Vahid Askarpour,[2] Chang Niu,[1] Adam Charnas,[1] Zhongxia Shang,[3] Yizhi Zhang,[3] Yaoqiao Hu,[4] Zhuocheng Zhang,[1] Pai-Ying Liao,[1] Kyeongjae Cho,[4] Haiyan Wang,[3] Mark Lundstrom,[1] Jesse Maassen,[2,*] Peide D. Ye[1,*]

[1]Elmore Family School of Electrical and Computer Engineering and Birck Nanotechnology Center, Purdue University, West Lafayette, IN 47907, United States

[2]Department of Physics and Atmospheric Science, Dalhousie University, Halifax, Nova Scotia B3H 4R2, Canada

[3]School of Materials Engineering, Purdue University, West Lafayette, IN 47907, United States

[4]Department of Materials Science and Engineering, The University of Texas at Dallas, Richardson, TX 75080, United States

*Address correspondence to: jmaassen@dal.ca (J.M.) and yep@purdue.edu (P.D.Y.)

**Abstract**

High drive current is a critical performance parameter in semiconductor devices for high-speed, low-power logic applications or high-efficiency, high-power, high-speed radio frequency (RF) analog applications. In this work, we demonstrate an $In_2O_3$ transistor grown by atomic layer deposition (ALD) at back-end-of-line (BEOL) compatible temperatures with a record high drain current exceeding 10 A/mm, the performance of which is 2-3 times better than all known transistors with semiconductor channels. A record high transconductance of 4 S/mm is also achieved among all transistors with a planar structure. It is found that a high carrier density and high electron velocity both contribute to this remarkably high on-state performance in ALD $In_2O_3$ transistors, which is made possible by the high-quality oxide/oxide interface, the metal-like charge-neutrality-level (CNL) alignment, and the high band velocities induced by the low density-of-state (DOS). Experimental Hall, I-V and split C-V measurements at room temperature confirm a high carrier density up to 6-7×$10^{13}$ /$cm^2$ and a high velocity of about $10^7$ cm/s. Ultra-thin oxide semiconductors, with a CNL located deep inside the conduction band, represent a promising new direction for the search of alternative channel materials for high-performance semiconductor devices.

Looking for a transistor with higher drain current ($I_D$) at low voltages has been a key pursuit in the development of modern semiconductor devices. High drain current is essential for high-speed, low-power logic applications for microelectronics and high-efficiency, high-power, high-speed applications for radio frequency (RF) power electronics. The criteria for achieving a transistor with high $I_D$ is rather simple because $I_D$ is determined by the product of carrier density ($n_{2D}$) and average velocity.[1] However, the demonstration of simultaneous high carrier density and high velocity has been difficult with the maximum $I_D$ ($I_{D,max}$) of reported semiconductor devices reaching 1-2 A/mm with Si[1], 3-4 A/mm with GaN[2] and 5 A/mm with graphene[3] (although graphene does not have a bandgap and therefore cannot be regarded as a semiconductor). The control of mobile carriers in conventional semiconductor devices is achieved by carrier accumulation at the oxide/semiconductor interface for enhancement-mode devices or carriers doped in the channel for depletion-mode devices. In this case, the maximum supported $n_{2D}$ is determined by $2E_{BD}\epsilon_{ox}\epsilon_0/q$, where $E_{BD}$ is the breakdown electric field of the gate oxide, $\epsilon_{ox}$ is the dielectric constant, $\epsilon_0$ is the vacuum permittivity, and q is the elementary charge. For example, $HfO_2$ has a typical breakdown electric field of 0.5 V/nm and dielectric constant of 15, leading to a maximum carrier density is about $8\times10^{13}$ /cm$^2$. However, in practical devices, such high carrier densities are difficult to achieve due to the U-shape distribution of interface traps.[4] As a result, the experimental $n_{2D}$ for actual inversion-mode Si devices is about $2\times10^{13}$ /cm$^2$, which is much lower than the ideal case. Moreover, the average carrier velocity can benefit from high $n_{2D}$ because the electrons sample higher energy states in the conduction band with increased velocities, which will be discussed later in this work. Such an effect is amplified in semiconductor materials with a single conduction band with low density of states (DOS).

Oxide semiconductor transistors are a mature technology in flat-panel display applications[5–7] and are being considered as promising high-performance devices for back-end-of-line (BEOL) compatible monolithic 3D integration.[8–11] An atomic layer deposition (ALD) based indium oxide ($In_2O_3$) transistor has been reported, featuring an atomically thin channel down to 0.5 nm and a high field-effect mobility ($\mu_{FE}$) over 100 cm$^2$/V·s, enabled by an atomically smooth surface and a charge neutrality level (CNL) located above the conduction band edge ($E_C$) for low resistance contacts.[12–17] The oxide insulator and oxide semiconductor interface was found to have a low interface trap density ($D_{it}$), resulting in a subthreshold slope (SS) down to 63.8 mV/dec at room temperature.[13] Note that although the SS of III-V compound semiconductor transistors can also be near 60 mV/dec, however, such low $D_{it}$ is located in their subthreshold regions. As a result, $n_{2D}$ is strongly affected by $D_{it}$ in the on-state due to the U-shape distribution of interface traps. Therefore, considering the CNL alignment above $E_C$ and the U-shape distribution of interface traps, an ALD $In_2O_3$ transistor is expected to have low $D_{it}$ in the on-state, leading to a high carrier density and a high electron velocity in scaled transistors.

In this work, we report an ALD $In_2O_3$ transistor with scaled channel length ($L_{ch}$) down to 7 nm achieving a high $I_{D,max}$ exceeding 10 A/mm, which is 2-3 times higher than the best known semiconducting materials such as Si, GaN, InGaAs, graphene, etc. A high transconductance ($g_m$) of 4 S/mm is also achieved, which is higher than all known transistors with a planar structure. This remarkably high on-state performance is attributed to the high carrier density up to 6-7×10$^{13}$ /cm$^2$ and the high velocity of about 10$^7$ cm/s, confirmed experimentally by Hall, I-V and split C-V measurements at room temperature. The simultaneous achievement of high density and high velocity, originating from a low DOS and high average electron energy, leads to the high near-ballistic current over 10 A/mm. The low contact resistance/resistivity induced by the unique CNL alignment of the metal/$In_2O_3$ interface in the conduction band and the wide bandgap of 3.0 eV for $In_2O_3$ being sustainable for ultra-scaled lateral devices are another two factors for remarkable high drain current. This work demonstrates how ultra-thin oxide semiconductors with a CNL located deep inside the conduction band can enable low contact resistance/resistivity, and highlights a new direction in the search of alternative channel materials for high-performance semiconductor devices.

Figure 1a illustrates the schematic diagram of the ALD $In_2O_3$ transistors in this work. The devices are comprised of a 3 nm-thick $HfO_2$ gate insulator, an ALD $In_2O_3$ channel with thickness ($T_{ch}$) from 2.5 nm to 3.5 nm, and Ni used as the gate electrode and the source/drain contacts. The detailed device fabrication is discussed in the Methods section. Figure 1b shows the high-angle annular dark field scanning transmission electron microscopy (HAADF-STEM) cross-sectional image with energy-dispersive X-ray spectroscopy (EDX) to highlight the Ni/In/Hf elements in a representative $In_2O_3$ transistor with $L_{ch}$ of 7 nm and $T_{ch}$ of 3.5 nm. Figure 1c and 1d present the $I_D$-$V_{GS}$ and $I_D$-$V_{DS}$ characteristics obtained by pulsed I-V measurements. A pulse width of 90 ns is used (unless otherwise specified) to suppress self-heating effects from the ultra-high drain current, so that the intrinsic electron transport properties of the devices can be probed, as described in the Methods section. Here, a high $I_{D,max}$ of 10.2 A/mm is achieved at $V_{GS}$=1.5 V and $V_{DS}$=1.4 V, which is 2-3 times higher than any best reported $I_{D,max}$ from devices with other semiconducting materials, including graphene. Figure 1e and 1f show the $I_D$-$V_{GS}$ on a linear scale and the $g_m$-$V_{GS}$ characteristics for a $V_{DS}$ of 1.4 V, both for the same device. A high $g_m$ of 4 S/mm for a $V_{DS}$ of 1.4 V is achieved, which is also higher than all known planar semiconductor devices. The $I_D$-$V_{GS}$, $I_D$-$V_{DS}$ and $g_m$-$V_{GS}$ characteristics of a device with the same gate stack and $L_{ch}$ of 100 nm are shown in Extended Data Fig. 1, which shows a high $I_{D,max}$ of 4.1 A/mm and a high $g_m$ of 1.3 S/mm even with this relatively long channel. The same characteristics of a long channel device with $L_{ch}$ of 1 µm are presented in Extended Data Fig. 2, showing clear current saturation. The remarkably high on-state performance is ascribed to the high electron density and high velocity of ALD $In_2O_3$, coupled with low contact resistance. Moreover, the wide bandgap of 3.0 eV in $In_2O_3$ makes it possible to sustain the high estimated lateral electric fields of ~ 2 MV/cm ($V_{DS}$=1.4 V and $L_{ch}$=7 nm). Similar electrical characterization of devices with $T_{ch}$ of 2.5 nm are also presented in Extended Data Fig. 3.

To understand the origin of the observed high current in ALD $In_2O_3$ transistors, the carrier density and velocity are characterized by Hall, I-V and split C-V measurements at room temperature. Figure 2a shows a photo image of the experimental Hall bar device, with a similar Ni/$HfO_2$/$In_2O_3$/Ni gate stack, having undergone the same thermal treatment process as the ALD $In_2O_3$ transistors. To minimize the impact of gate leakage current, a thickness of 16 nm $HfO_2$ is used in the Hall measurements. Figure 1b shows the $R_{xy}$ versus B field characteristics with $T_{ch}$ of 2.5 nm. A $n_{2D}$ of $5.4 \times 10^{13}$ /cm$^2$ and Hall mobility ($\mu_{Hall}$) of 48.2 cm$^2$/V·s are achieved. Figure 2c shows the $V_{GS}$-dependent $n_{2D}$ and $\mu_{Hall}$. A maximum $n_{2D}$ of $7 \times 10^{13}$ /cm$^2$ is obtained, which can be depleted or modulated by gate voltage, and is significantly higher than other semiconducting materials such as Si, GaN, etc. The high $n_{2D}$ is very close to the maximum charge density supported by $HfO_2$ as calculated above, indicating a high-quality oxide/oxide semiconductor interface without significant trapped charges. Extended Data Fig. 5 presents a typical $I_D$-$V_{GS}$ curve of an ALD $In_2O_3$ transistor with $T_{ch}$ of 1.2 nm, $L_{ch}$ of 800 nm and 5 nm $HfO_2$ thickness, showing a steep subthreshold slope (SS) of 63.5 mV/dec corresponding to a $D_{it}$ of $6 \times 10^{11}$ cm$^{-1}$·eV$^{-1}$, suggesting a high-quality $HfO_2$/$In_2O_3$ interface. The SS in Fig. 1 is much larger due to the relatively thick channel designed to achieve high drain current and short-channel effects, but the $D_{it}$ at the HfO$_2$/In$_2$O$_3$ interface is expected to be similar and independent of channel thickness because of the similar atomic configuration at the interface. More importantly, the maximum charge density is not limited by the high D$_{it}$ in the tail of the U-shape distribution as with conventional metal-oxide-semiconductor (MOS) devices like InGaAs. This is because In$_2$O$_3$ has a CNL about 0.4 eV above the E$_C$, so that D$_{it}$ in the on-state is expected to be even lower than in the subthreshold region. Thus, with ALD In$_2$O$_3$ transistors containing an oxide/oxide interface, the capability of carrier density modulation by gate oxide is much higher than those with a conventional III-V oxide/semiconductor interface. The high carrier density is also further verified by split C-V measurements as shown in Extended Data Fig. 6. I$_D$ can be calculated simply by the product of charge density and average velocity. Because of the ultra-scaled channel length of 7 nm compared to the estimated mean free path for backscattering of roughly 8 nm in In$_2$O$_3$ (see Extended Data Fig. 4), the average velocity is estimated to be the ballistic injection velocity. Using the transfer characteristics in Fig. 1c for I$_D$ and the Hall measurements in Fig. 2c for n$_{2D}$, a high electron velocity in the range 0.85-1.05×10$^7$ cm/s is obtained at high n$_{2D}$ and shown in Fig. 2d. This electron velocity might still be under-estimated because non-ideal effects, such as contact resistance, short channel effects etc., are not considered, suggesting that the electron injection velocity in ALD In$_2$O$_3$ transistors can be even higher. Such high carrier density and velocity have contributed to a high I$_{D,max}$ over 10 A/mm, which is significantly higher than other state-of-the-art semiconductor devices.

To elucidate the nature of the high current in In$_2$O$_3$, density functional theory (DFT) calculations of the electronic properties were conducted (see Methods section for details). Figure 3b shows the band structure of bulk In$_2$O$_3$ and two In$_2$O$_3$ 2D slabs with thicknesses of 1.98 nm and 0.95 nm; the corresponding atomic structures are presented in Fig. 3a. In all cases there is a single nearly-isotropic conduction band located at the zone center, comprised of mostly In s-orbital character, with effective masses of 0.17 $m_0$ (bulk), 0.23 $m_0$ (1.98 nm slab) and 0.30 $m_0$ (0.95 nm slab). A key feature of the band structure is the absence of higher-energy secondary bands until well above 0.5 eV relative to the conduction band edge. Figure 3c shows the DOS of bulk $In_2O_3$ with an assumed thickness of 3.5 nm, the two $In_2O_3$ slabs, and for comparison bulk Si with an assumed thickness of 3.5 nm. $In_2O_3$ displays a much smaller DOS than Si, due to its lower effective mass and single conduction band (the DOS effective mass of bulk $In_2O_3$ is 0.17 $m_0$ compared to 1.06 $m_0$ for bulk Si). Thus, for a given $n_{2D}$, the Fermi level is located deeper inside the conduction band of $In_2O_3$ compared to Si. For example, a carrier density of $5\times10^{13}$ /cm$^2$ corresponds to a Fermi level of roughly 0.5 eV and 0.1 eV above the band edge for bulk $In_2O_3$ and Si, respectively, as seen in Extended Data Fig. 7. Moreover, from Extended Data Fig. 7 one finds that the smaller effective mass of $In_2O_3$ results in a higher average band velocity at a given energy. From the DFT band structure the ideal ballistic current is computed versus $n_{2D}$ and shown in Fig. 3d. This current represents an ideal upper limit for a short-channel device in which transport is ballistic, the current is controlled by injection from the source contact only ($V_{DS} \gg k_BT/q$), and there is no potential barrier. With $In_2O_3$ the minimum carrier concentration needed to achieve a current of 10 A/mm is $2\times10^{13}$ /cm$^2$, while with Si a carrier concentration near $5\times10^{13}$ /cm$^2$ is required. To understand why $In_2O_3$ can carry a larger current than Si, the ballistic injection velocity is obtained from the ratio of the ideal ballistic current over the charge density, which is plotted in Fig. 3e. The velocities of $In_2O_3$ are significantly higher than Si, particularly for large $n_{2D}$. For $5\times10^{13}$ /cm$^2$, $In_2O_3$ shows velocities ranging from $3.4\times10^7$ cm/s (0.95 nm slab) to $3.7\times10^7$ cm/s (1.98 nm slab) compared to $1.4\times10^7$ cm/s with Si. The origin of this 2-3 times velocity enhancement is traced back to $In_2O_3$ having a single conduction band with relatively low effective mass, leading to a combination of higher band velocities and a Fermi level located deep in the band where electrons sample higher energy states. $In_2O_3$ also benefits from its secondary bands being high enough in energy to not play a significant role in transport, which would otherwise lower the average electron velocity and energy.

Therefore, the two major contributors to the high $I_{D,max}$ in ALD $In_2O_3$ transistors are high carrier density and high velocity, with the former originating from the high-quality oxide/oxide interface and a metal-like CNL alignment and the latter arising from the low DOS and high-carrier-induced velocity enhancement. Moreover, the ALD $In_2O_3$ transistors are aided by low contact resistances and the wide bandgap of $In_2O_3$ for laterally scaled channel lengths. The above first-principles calculations are based on a single crystal structure. The crystal structure of the experimental devices is confirmed to be polycrystalline. As shown in the TEM cross-sectional image and X-Ray Diffraction (XRD) data in Extended Data Fig. 8, unannealed ALD $In_2O_3$ is amorphous while annealed ALD $In_2O_3$ at 275 °C in $O_2$ for 60 s, the same process condition for device fabrication, is polycrystalline. It would be natural to assume that the charge transport has much more scatterings in the polycrystalline boundaries and that carrier mobility (velocity) might be reduced significantly. However, our DFT calculations on polycrystalline $In_2O_3$ reveals that there is a possible mobility enhancement effect such that polycrystalline and single crystalline $In_2O_3$ could have similar charge transport behavior and comparable electron mobility and velocity, as supported by Extended Data Fig. 9.

Figure 4a and 4b present the benchmark of $I_{D,max}$ and $g_m$ versus channel length of selected high performance semiconductor devices with different channel materials. ALD $In_2O_3$ displays a $I_{D,max}$ that is 2-3 times better than the best reported transistors with different channel materials including Si, InGaAs, GaN, graphene etc., and also demonstrates the best $g_m$ compared to all other semiconductor materials with a planar device structure, due to the simultaneous high carrier density and high velocity. Data used in this work is summarized in Extended Data Table I.[2,3,8,11,18–28]

In summary, an ALD $In_2O_3$ transistor with a record high $I_{D,max}$ exceeding 10 A/mm is reported, which is 2-3 times higher than that of all known semiconductor devices. It is found that high carrier densities, resulting from the high-quality oxide/oxide interface and a metal-like CNL alignment, and high carrier velocities, originating from the low DOS and high-carrier-induced velocity enhancement, are the two major contributors to the remarkably high $I_{D,max}$ in ALD $In_2O_3$ transistors. Hall, I-V and split C-V measurements at room temperature confirm a high carrier density up to $6-8\times10^{13}$ /cm$^2$ and a high velocity of about $10^7$ cm/s. Ultra-thin oxide semiconductors with a CNL located deep inside the band represent a new route in the search of alternative channel materials for high-performance semiconductor devices.

**Methods**

**Device Fabrication.** The device fabrication process started with solvent cleaning of the p+ Si substrate with thermally grown 90 nm $SiO_2$. A bi-layer lift-off process was then adopted for the sharp lift-off of 30 nm Ni buried metal gate formed by e-beam evaporation, utilizing photoresist PMGI SF9 + AZ1518 stack, patterned by Heidelberg MLA150 Maskless Aligner. This step is critical to avoid sidewall metal coverage and to enable a high-quality ultra-scaled ALD gate dielectric and semiconducting channel. $HfO_2$ of 3.0 nm as a gate dielectric was grown by ALD at 200°C using $[(CH_3)_2N]_4Hf$ (TDMAHf) and $H_2O$ as Hf and O precursors, respectively. Then, $In_2O_3$ thin films with various thicknesses were grown by ALD at 225°C using $(CH_3)_3In$ (TMIn) and $H_2O$ as In and O precursors, respectively. $N_2$ is used as the carrier gas at a flow rate of 40 sccm. The base pressure is 169 and 437 mTorr at a $N_2$ flow rate of 0 sccm and 40 sccm, respectively. Concentrated hydrochloric acid was applied for the channel isolation. S/D ohmic contacts were formed by e-beam evaporation of 30 nm Ni in two steps to avoid the difficulty of the sub-10 nm lift-off process, as employed before. A two-step e-beam lithography process was performed by the formation of first the source electrode and then the drain electrode, utilizing diluted e-beam resist AR-P 6200 (2.5%), patterned by a JEOL JBX-8100FS E-Beam Writer. Then, inductively coupled plasma (ICP) dry etching using $BCl_3$/Ar plasma was applied to accurately define the channel width. The devices were annealed in $O_2$ at 275 °C for 1 min to further improve the performance.

**Material Characterization.** The cross-sectional TEM specimens were fabricated by focused ion beam (FIB) using a FEI Helios G4 UX dual beam scanning electron microscope (SEM). Before FIB milling, a platinum (Pt) layer was deposited to protect the surface from ion damage. To remove FIB induced damages, the TEM samples were ion polished using a low current (27 pA) at a low voltage (2 keV) multiple times during the final steps.

Transmission electron microscopy (TEM) analyses and energy-dispersive X-ray spectroscopy (EDX) element mapping were carried out on a Thermo-Fischer FEI Talos 200X TEM microscope operated at 200 kV equipped with a high angle annular dark field (HAADF) detector and SuperX EDS with four silicon drift detectors. All the EDX maps were captured under the drift correction mode.

**Device Characterization.** The DC characterization is performed using a B1500A Semiconductor Device Analyzer. The pulse characterization is performed using a B1530A Waveform Generator/Fast Measurement Unit (WGFMU). The device is grounded at the source, while the drain and gate are connected to two channels of the WGFMU via Remote Sense/Switch Units (RSU). A waveform with a pulse delay time (PD) of 2 μs, a rise time (RT) of 60 ns, a pulse width (PW) of 90 ns and a fall time (FT) of 60 ns is adopted to maximize device reliability while minimizing pulse-induced AC response during measurements. Two channels carry the waveform simultaneously with a chronological difference less than 1 ns, confirmed with TDS5000 Digital Phosphor Oscilloscopes. Data is collected during an averaging time (AT) of 50 ns after a measurement delay (MD) of 2.1 μs to wait for the signal to stabilize. No bias is applied at each channel during the pulse interval to dissipate generated heat. During output characterization, the amplitude of the $V_{DS}$ pulse increases in 100 mV steps while the amplitude of the $V_{GS}$ pulse remains constant. During transfer characterization, the amplitude of the $V_{DS}$ pulse increases in 50 mV steps while the amplitude of the $V_{GS}$ pulse remains constant. The transfer data is spliced with the on-state data from the pulse characterization while the off-state data comes from the DC characterization, due to the limited signal sensing range of the B1530A WGFMU in the high-current mode (100 μA-10 mA) and the relatively low resolution of the RSUs.

Pulse characterization is critical to alleviate severe self-heating-induced burning and minimize time-dependent dielectric breakdown of the ultra-thin gate insulator. The pulse data is in line with the DC data in the low power range. While DC characterization inevitably causes severe self-heating and resulting device failure, pulse characterization enables repeatable measurements. The details of pulsed I-V measurements and the impact of self-heating effects are shown in Extended Data Fig. 10.

**Theoretical and computational details.** The density functional theory (DFT) calculations were carried out using the Quantum Espresso (QE) code[29,30]. Optimized norm-conserving Vanderbilt pseudopotentials[31] and the Perdew-Burke-Ernzerhof (PBE)[32] exchange-correlation were adopted. A kinetic energy cut-off of 140 Ry and a Monkhorst-Pack[33] Brillouin zone sampling of 6×6×6 for bulk (primitive cell) $In_2O_3$ and 3×3×1 for the 2D slabs of $In_2O_3$ were used. Crystalline $In_2O_3$ is body-centered cubic with 40 atoms in the primitive cell.[34] The 80-atom conventional cubic cell, was used to generate the 0.95 nm $In_2O_3$ slab. The indium layer at the boundary of the cell was removed and the top and bottom oxygen layers were passivated with hydrogen atoms (one hydrogen per oxygen). The same process was used to generate the 1.98 nm slab, but starting with a 1×1×2 supercell of the bulk cubic cell. A vacuum layer of 25 Å was added to both slabs. With bulk silicon an energy cut-off of 120 Ry and a Brillouin zone sampling of 16×16×16 (primitive cell) was employed. All atoms and lattice vectors were fully relaxed, with energies and forces converged to $10^{-5}$ Ry and $10^{-4}$ Ry/Bohr, respectively. Spin-orbit coupling was not included.[35] The optimized lattice constant of bulk $In_2O_3$ is 10.30 Å, which is consistent with other studies.[34-37] The in-plane lattice constants for the 1.98 nm slab are 10.28 Å and 10.32 Å, and those for the 0.95 nm slab are 10.26 Å and 10.36 Å. Bulk Si has an optimized lattice constant of 5.47 Å, which is close to the measured value of 5.43 Å.[38] The calculated band gaps for bulk $In_2O_3$, the 1.98 nm slab, the 0.95 nm slab, and bulk Si are 0.94 eV, 1.27 eV, 1.88 eV, and 0.61 eV, respectively. The band gaps of bulk $In_2O_3$ and Si are below the experimental values of 2.7-3.25 eV[37,39-42] and 1.12 eV[43], however this study focuses only the conduction states. The effective mass of bulk $In_2O_3$, equal to 0.17 $m_0$, is consistent with some reported values.[36-37,44-46]

The Wannier code[47] was employed to generate maximally localized Wannier functions, which were used to calculate electron energies and velocities on the following $k$-grids: 200×200×200 for bulk $In_2O_3$ and bulk Si, and 200×200×1 for the $In_2O_3$ slabs. Calculated quantities include the density of states (DOS), $D(E) = 2\sum_{k,n}\delta(E - E_{k,n})/\Omega$, the distribution of modes (DOM), $M(E) = (h/2)\sum_{k,n}|v_x(k,n)|\delta(E - E_{k,n})/\Omega$, and the average velocity along the transport direction (the $x$-direction), $v_x^+(E) = \sum_{k,n}|v_x(k,n)|\delta(E - E_{k,n})/\sum_{k,n}\delta(E - E_{k,n})$[48], where $E_{k,n}$ are the electron eigen-energies for a given wavevector $k$ and band index $n$, and $v_x(k) = (1/\hbar)\partial E_k/\partial k_x$. The carrier concentration and ideal ballistic current are calculated using $n_{2D} = (1/2)\int_{E_c}^{\infty} D(E)f_0(E, E_F)dE$ and $I_{ball} = (2q/h)\int_{E_c}^{\infty} M(E)f_0(E, E_F)dE$[49], where $f_0(E, E_F)$ is the Fermi-Dirac distribution and $E_F$ is the Fermi energy (the factor of two in $n_{2D}$ accounts for only half the states being occupied in an ideal ballistic device[50]). The ballistic injection velocity is obtained from $v_{inj} = I_{ball}/(qn_{2D})$. From the experimental Hall data, the average mean free path for backscattering is fitted by: 1) finding the location of the Fermi level for which the calculated and measured $n_{2D}$ are the same, 2) calculating the average distribution of modes at that Fermi level, $\langle M \rangle = \int_{E_c}^{\infty} M(E)[-\partial f_0/\partial E]dE$, and 3) obtaining the average mean free path for backscattering using the relation $\langle \lambda \rangle = (h/2q)n_{2D}\mu_{Hall}/\langle M \rangle$.

**Acknowledgements**

This work was supported in part by the Semiconductor Research Corporation (SRC) nCore Innovative Materials and Processes for Accelerated Compute Technologies (IMPACT) Center, in part by the Air Force Office of Scientific Research (AFOSR), and in part by SRC/Defense Advanced Research Projects Agency (DARPA) Joint University Microelectronics Program (JUMP) Applications and Systems-driven Center for Energy Efficient integrated Nano Technologies (ASCENT) Center. V.A. and J.M. acknowledge support from NSERC and Compute Canada.

**Author Contributions**

P.D.Y. conceived the idea and proposed research. M.S. and Z.L. developed the ALD $In_2O_3$ thin film deposition and device fabrication process. Z.L. did the device fabrication, electrical measurements, and analysis on thickness and EOT scaling of ALD $In_2O_3$ devices. A.C. did the split C-V measurements and XRD characterization. V.A. and J.M. performed the first-principles calculations and analysis. Y.H. and K.C. provided the mobility model on poly-crystal. M.L. provided insights on ballistic transistors and understanding of device physics. Z.S, Y.Z and Z.L performed the STEM and EDX measurements. M.S., Z.L, V.A., J.M. and P.D.Y. co-wrote the manuscript and all authors commented on it.

**Financial Interest Statement**

The authors declare no competing financial interest.

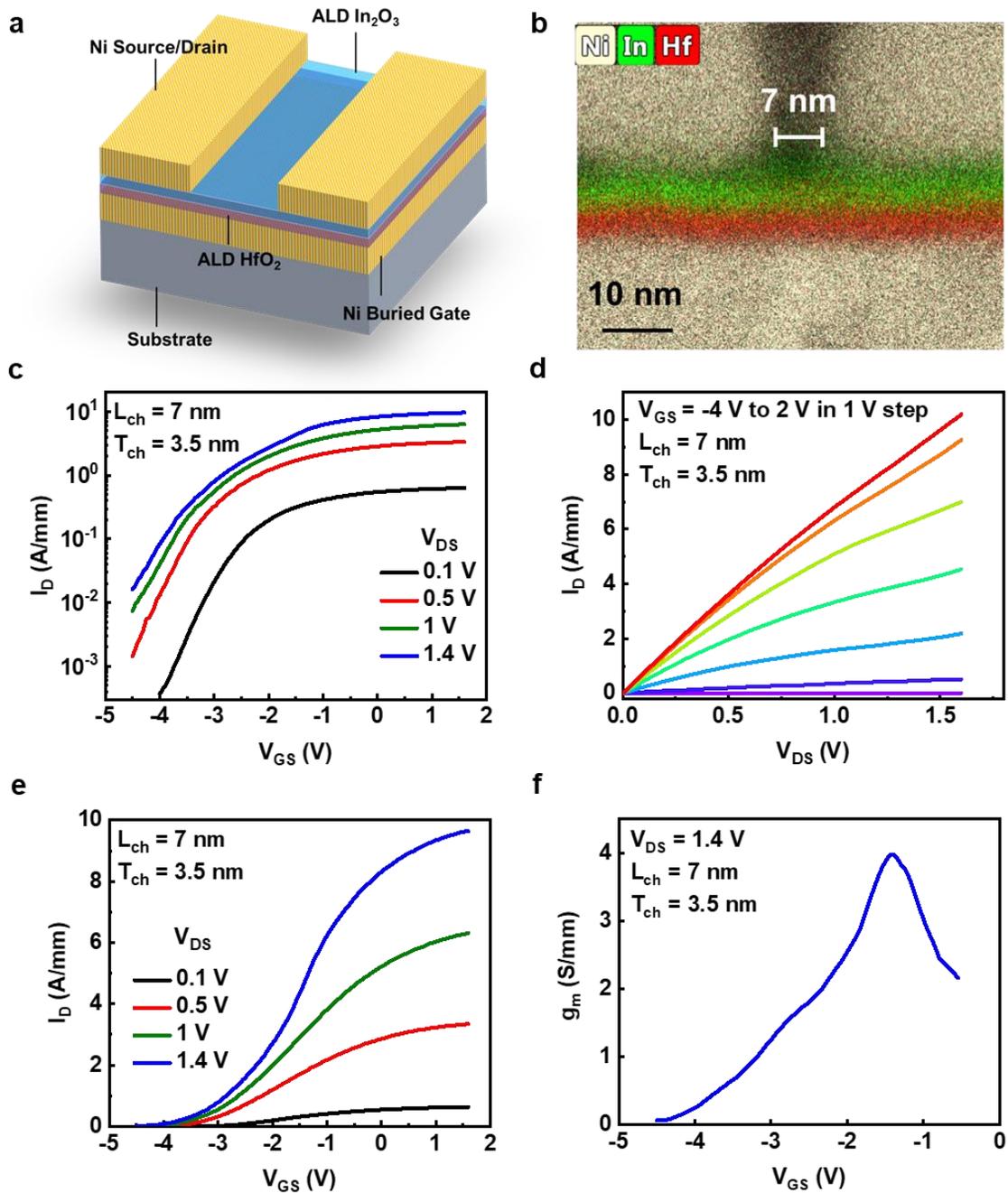

**Fig. 1 | Performance of scaled ALD In₂O₃ transistors. a,** Schematic diagram of an ALD In$_2$O$_3$ transistor. **b,** HAADF-STEM cross sectional image with EDX element mapping of an In$_2$O$_3$ transistor with $L_{ch}$ of 7 nm, $T_{ch}$ of 3.5 nm and 3 nm HfO$_2$ as gate insulator. **c,** $I_D$-$V_{GS}$ in log scale and **d,** $I_D$-$V_{DS}$ characteristics of a representative ALD In$_2$O$_3$ transistor with $L_{ch}$ of 7 nm, $T_{ch}$ of 3.5

nm. The device exhibits high $I_D$ exceeding 10 A/mm. **e,** $I_D$-$V_{GS}$ characteristics in linear scale and **f,** $g_m$-$V_{GS}$ characteristics at $V_{DS}$ of 1.4 V of the same device, showing a maximum $g_m$ of 4 S/mm. From Fig. 1d and Fig. 1e, $n_{2D}$ is estimated to be ~ $7.6 \times 10^{13}$/cm$^2$ using $n_{2D}=C_G(V_{GS}-V_T)$. Details are seen in Extended Data Fig. 2.

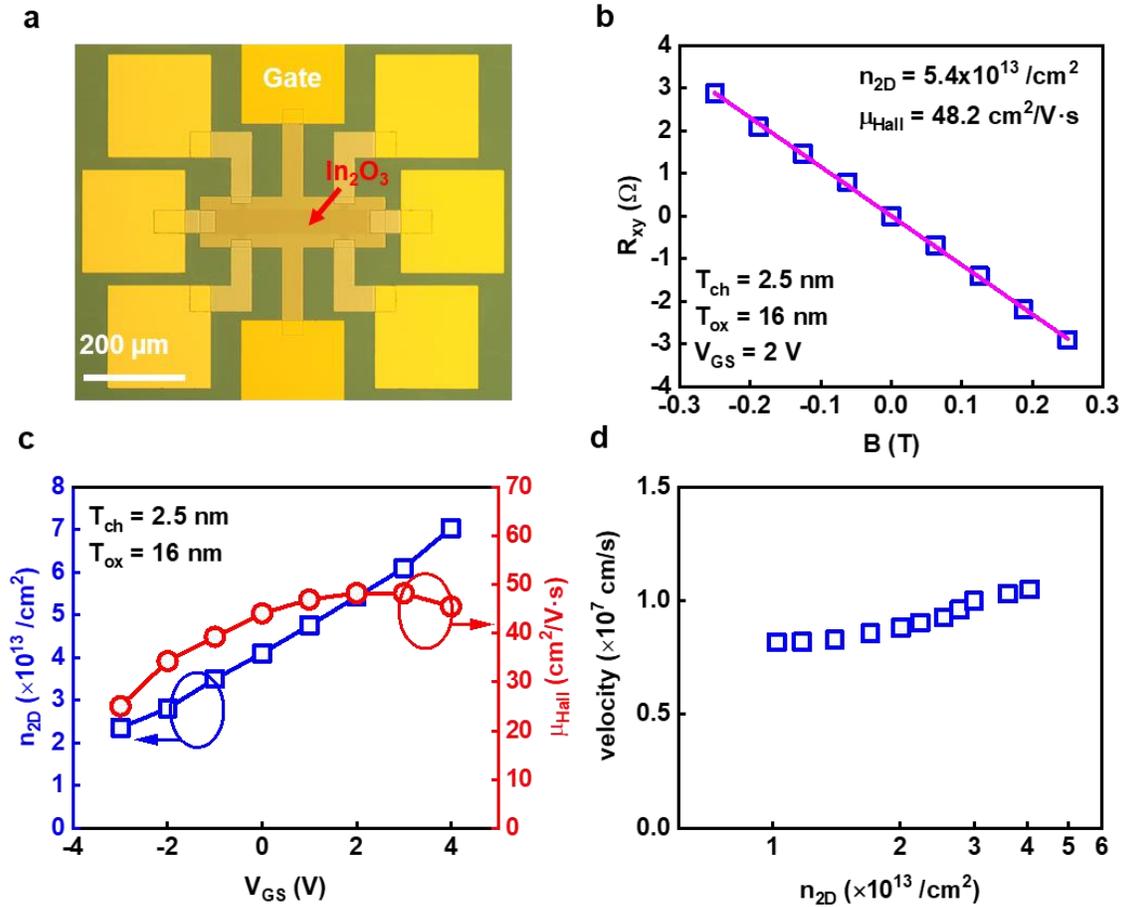

**Fig. 2 | Carrier density and velocity in ultra-thin ALD In$_2$O$_3$. a,** Photo image of a Hall bar device with back gate structure with 16 nm HfO$_2$ as gate insulator and 2.5 nm In$_2$O$_3$ as channel. Hall devices underwent the same thermal treatment process. A thick oxide is used here to avoid the impact of gate leakage current. **b,** Hall measurement of R$_{xy}$ versus B field with the gate electrode at 2 V. n$_{2D}$ of 5.4×10$^{13}$ /cm$^2$ and µ$_{Hall}$ of 48.2 cm$^2$/V·s are achieved. **c,** n$_{2D}$ and µ$_{Hall}$ versus V$_{GS}$ characteristics for V$_{GS}$ from -3 V to 4 V. **d,** Velocity versus n$_{2D}$ extracted from the I$_D$-V$_{GS}$ characteristics in Fig. 1 and carrier density from Hall measurements.

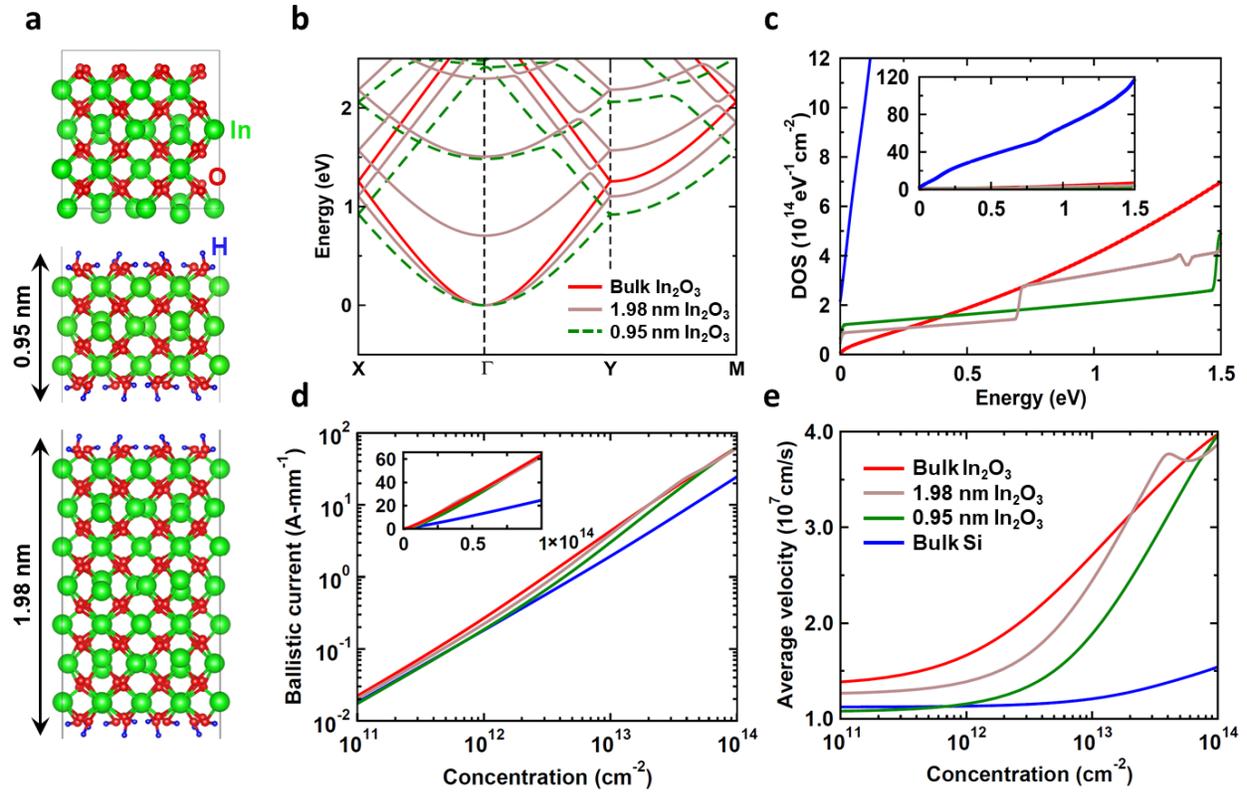

**Fig. 3 | First-principles simulations of In$_2$O$_3$. a,** Atomic structure of bulk In$_2$O$_3$ (top), 0.95 nm-thick In$_2$O$_3$ (middle) and 1.98 nm-thick In$_2$O$_3$ (bottom). **b,** Calculated band structure of bulk and nanoscale In$_2$O$_3$. **c,** Comparison of DOS versus electron energy (relative to the band edge) between In$_2$O$_3$ and Si. In$_2$O$_3$ presents a much lower DOS such that for a given carrier density electrons in In$_2$O$_3$ have a much higher energy than those in Si. **d,** Ideal ballistic current versus electron concentration. At high concentrations In$_2$O$_3$ can carry significantly more current than Si. **e,** Average (ballistic injection) velocity versus electron concentration in In$_2$O$_3$ and Si. In$_2$O$_3$ presents higher average velocities because of its lower DOS and higher band velocities. For panels **c** and **d**, the properties of bulk In$_2$O$_3$ and bulk Si are obtained assuming a thickness of 3.5 nm.

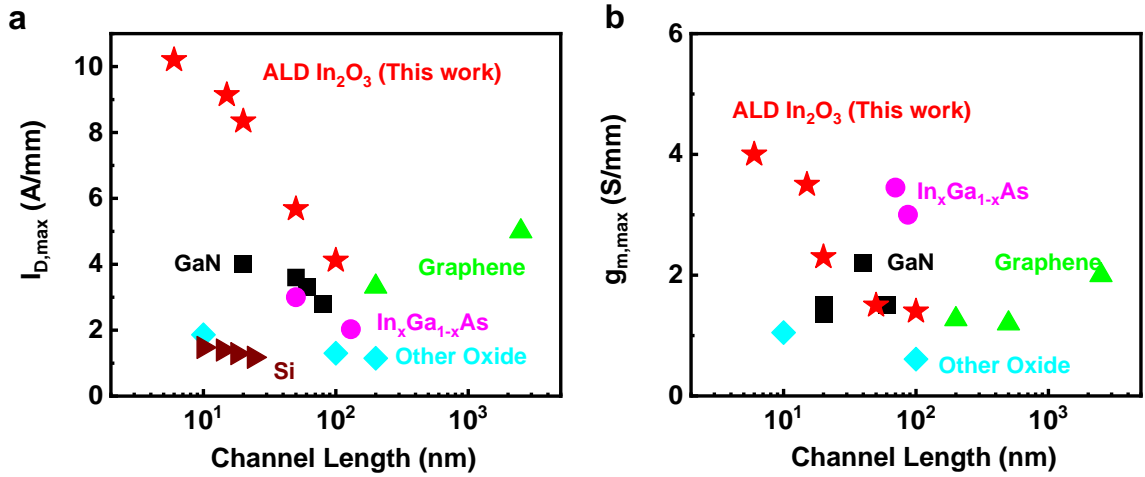

**Fig. 4 | Benchmarking of ALD In₂O₃ with other semiconductor materials.** Comparison of **a,** $I_{D,max}$ and **b,** $g_m$ versus channel length characteristics with other high-performance semiconductor materials and graphene. Transistors with ALD In₂O₃ demonstrate the largest $I_D$ and $g_m$ compared to all known semiconductors with a planar structure due to the high carrier density and high velocity.

# Supplementary Information

# A Nanometer-Thick Oxide Semiconductor Transistor with Ultra-high Drain Current

Zehao Lin,[1] Mengwei Si,[1] Vahid Askarpour,[2] Chang Niu,[1] Adam Charnas,[1] Zhongxia Shang,[3] Yizhi Zhang,[3] Yaoqiao Hu,[4] Zhuocheng Zhang,[1] Pai-Ying Liao,[1] Kyeongjae Cho,[4] Haiyan Wang,[3] Mark Lundstrom,[1] Jesse Maassen,[2,*] Peide D. Ye[1,*]

[1]Elmore Family School of Electrical and Computer Engineering and Birck Nanotechnology Center, Purdue University, West Lafayette, IN 47907, United States

[2]Department of Physics and Atmospheric Science, Dalhousie University, Halifax, Nova Scotia B3H 4R2, Canada

[3]School of Materials Engineering, Purdue University, West Lafayette, IN 47907, United States

[4]Department of Materials Science and Engineering, The University of Texas at Dallas, Richardson, TX 75080, United States

*Address correspondence to: jmaassen@dal.ca (J.M.) and yep@purdue.edu (P.D.Y.)

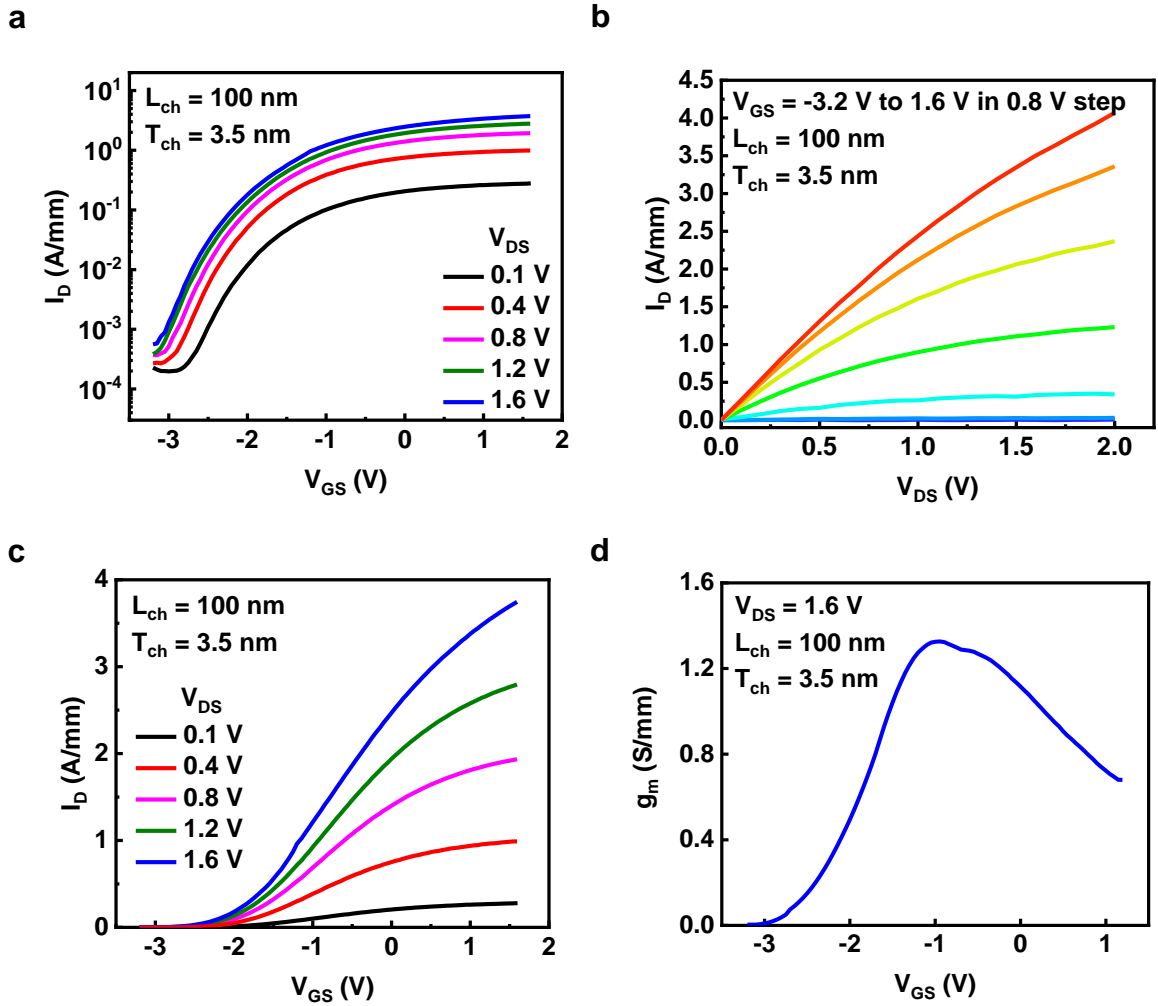

**Extended Data Fig. 1 | Performance of ALD In$_2$O$_3$ transistors with 100 nm channel length. a,** I$_D$-V$_{GS}$ in log scale and **b,** I$_D$-V$_{DS}$ characteristics of a representative ALD In$_2$O$_3$ transistor with L$_{ch}$ of 100 nm, T$_{ch}$ of 3.5 nm, showing I$_{D,max}$ of 4.1 A/mm. **c,** I$_D$-V$_{GS}$ characteristics in linear scale and **d,** g$_m$-V$_{GS}$ characteristics at a V$_{DS}$ of 1.6 V for the same device, showing maximum g$_m$ of 1.3 S/mm.

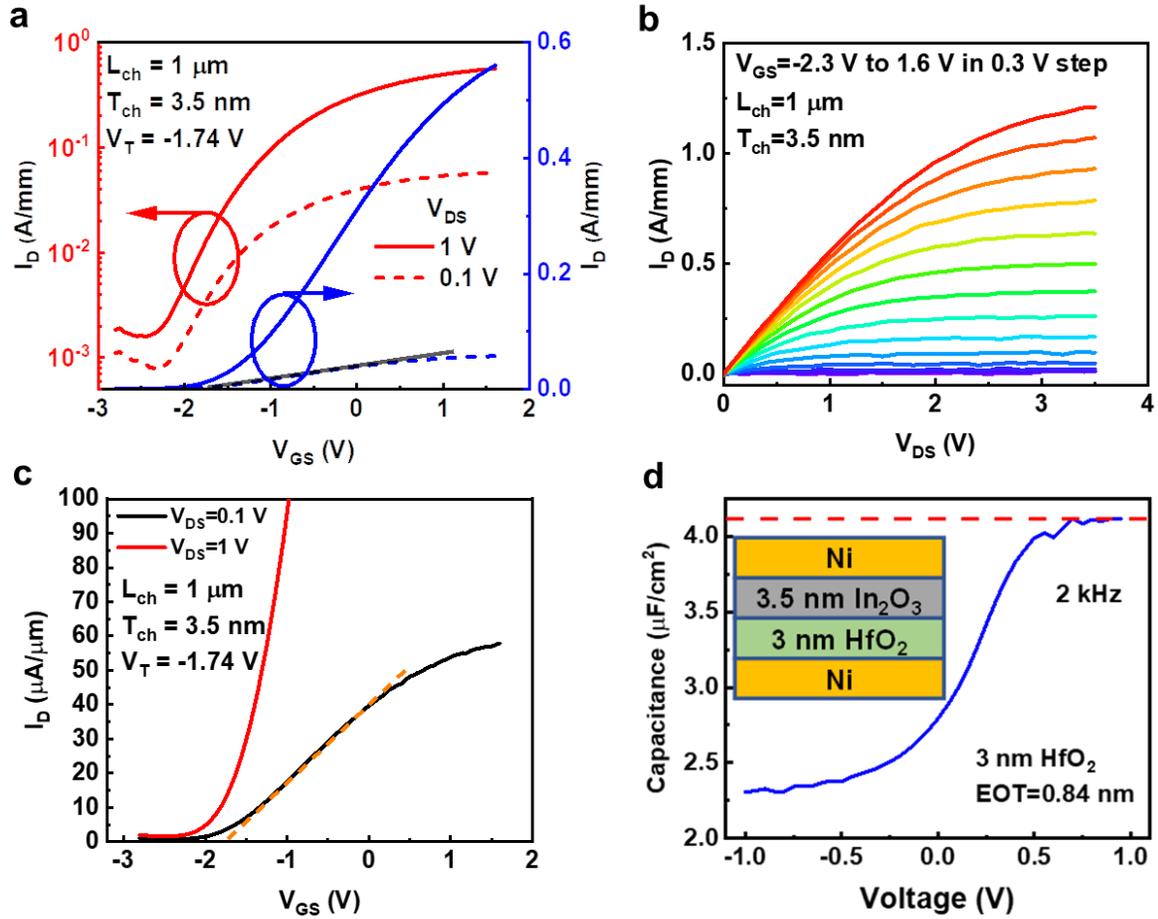

**Extended Data Fig. 2 | Performance of ALD In$_2$O$_3$ transistors with 1 μm channel length.**

**a,** I$_D$-V$_{GS}$ transfer characteristics in log and linear scales and **b,** I$_D$-V$_{DS}$ output characteristics of a representative ALD In$_2$O$_3$ transistor with L$_{ch}$ of 1 μm and T$_{ch}$ of 3.5 nm, showing clear saturation I$_{D,max}$ of 1.25 A/mm at V$_{DS}$=3.5 V. **c,** I$_D$-V$_{GS}$ characteristics in linear scale at sub-threshold region showing V$_T$ of -1.7 V determined by linear extrapolation and **d,** C-V measurement of the gate stack capacitor (Ni/3 nm HfO$_2$/3.5 nm In$_2$O$_3$/Ni) at 2 kHz, fabricated together with the ALD In$_2$O$_3$ transistor on the same chip. The n$_{2D}$ under V$_{GS}$=1.5 V for a transistor of L$_{CH}$=7 nm, as shown in Fig. 1d and Fig. 1e, is estimated to be ~ 7.6×10$^{13}$/cm$^2$ using n$_{2D}$=C$_G$(V$_{GS}$-V$_T$). The gate capacitance is given by C$_G$=C$_{ox}$C$_S$/(C$_{ox}$+C$_S$), where the semiconductor or quantum capacitance C$_S$ of In$_2$O$_3$ is about 5-8 times larger than $C_{ox}$ of 4.4 µF/cm$^2$. $C_S$ is also calculated in Extended Data Fig. 4(b). $V_T$ must be determined from a long channel device.

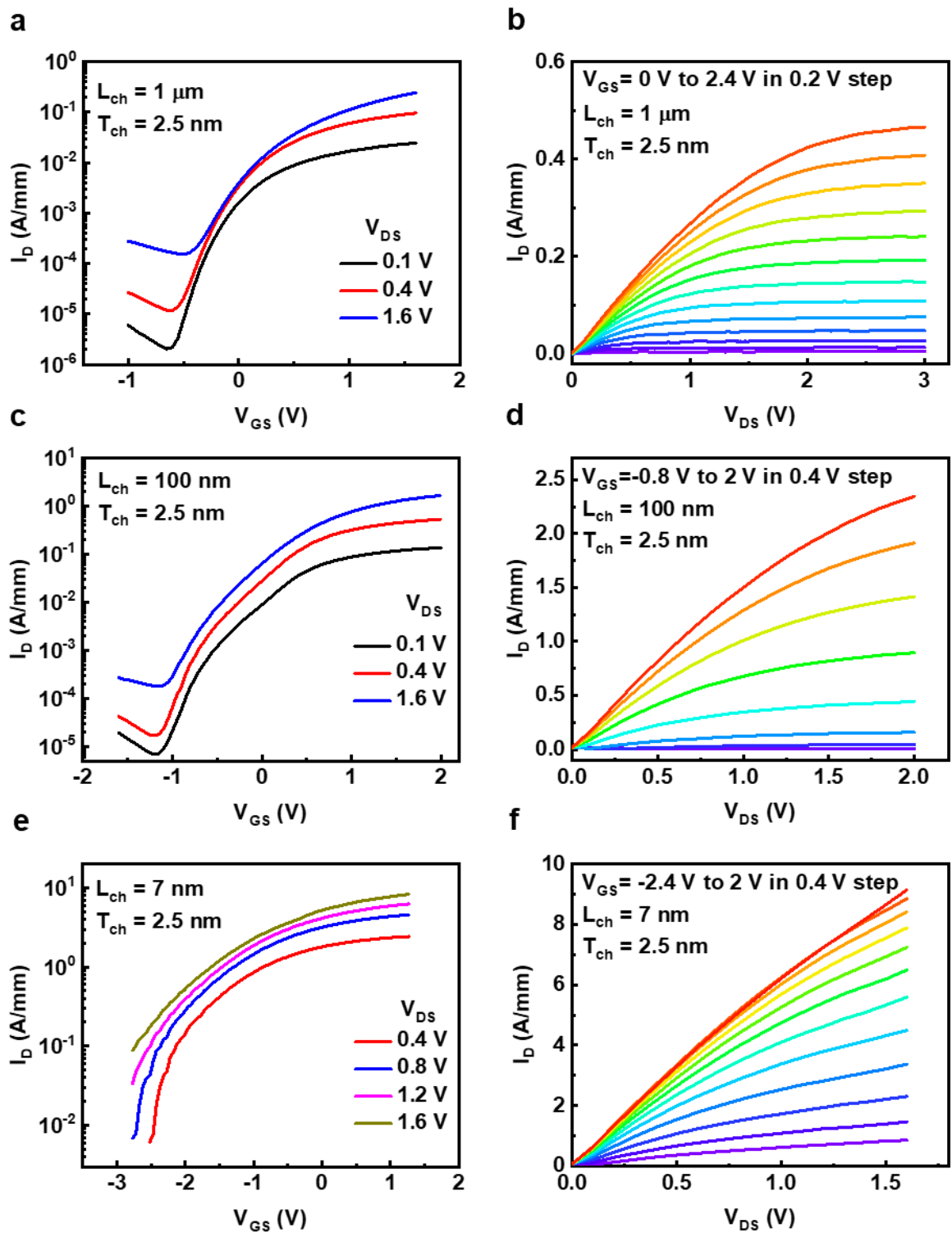

**Extended Data Fig. 3 | Performance of ALD In$_2$O$_3$ transistors with 2.5 nm channel thickness.**

**a, c, e,** I$_D$-V$_{GS}$ transfer characteristics in log scale and **b, d, f,** I$_D$-V$_{DS}$ output characteristics of representative ALD $In_2O_3$ transistors with $T_{ch}$ of 2.5 nm and $L_{ch}$ of 1 µm, 100nm and 7 nm, respectively. From the $I_D$-$V_{GS}$ characteristics on the linear scale in the sub-threshold region, $V_T$ is determined to be ~ 0 by linear extrapolation in long channel devices as shown in **a** and **b**. The long channel device is an enhancement-mode device. The $n_{2D}$ under $V_{GS}$=2.4 V is estimated to be ~ $6.0 \times 10^{13}$/cm$^2$ using $n_{2D}=C_G(V_{GS}-V_T)$, which is in good agreement with the results obtained from Hall measurements presented in Fig. 2. Here, $C_G=C_{ox}C_S/(C_{ox}+C_S)$, where the semiconductor or quantum capacitance $C_S$ of $In_2O_3$ is about 5-8 times larger than $C_{ox}$ of 4.4 µF/cm$^2$. $C_S$ is also calculated in Extended Data Fig. 4(b). $V_T$ must be obtained from a long channel device.

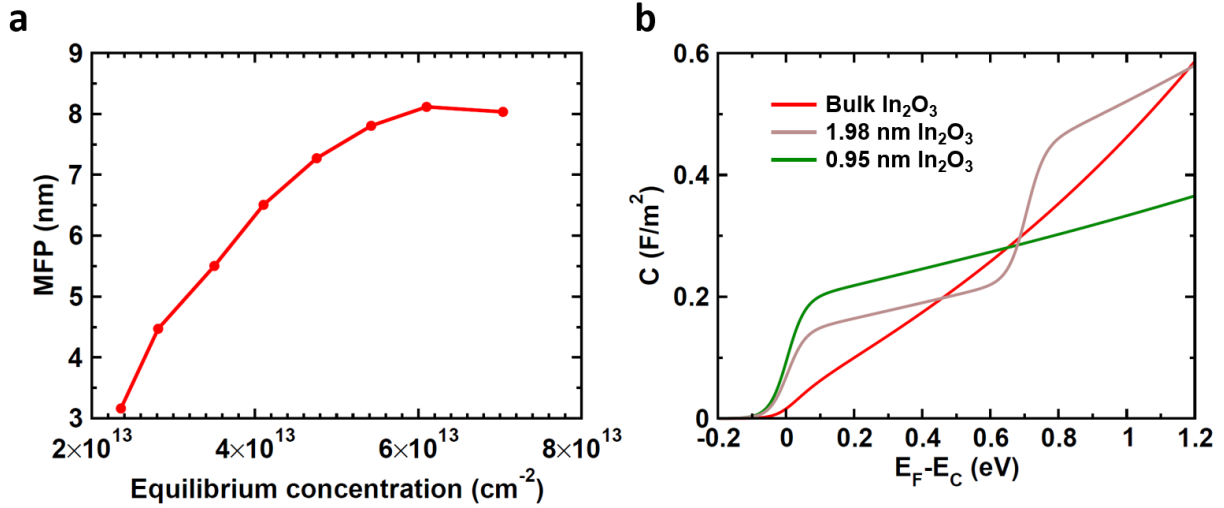

**Extended Data Fig. 4 | Calculated mean free path and capacitance of $In_2O_3$. a**, Average mean free path (MFP) for backscattering extracted from Hall measurements and DFT calculations versus equilibrium electron concentration. The MFP for backscattering is fitted to the properties of bulk $In_2O_3$ with a thickness of 2.5 nm. **b**, Calculated semiconductor capacitance of $In_2O_3$, approximated as $C_s \approx q\partial n_{2D}/\partial(E_F/q)$, versus Fermi level relative to the conduction band edge. The capacitance of bulk $In_2O_3$ is converted to 2D units by assuming a thickness of 2.5 nm.

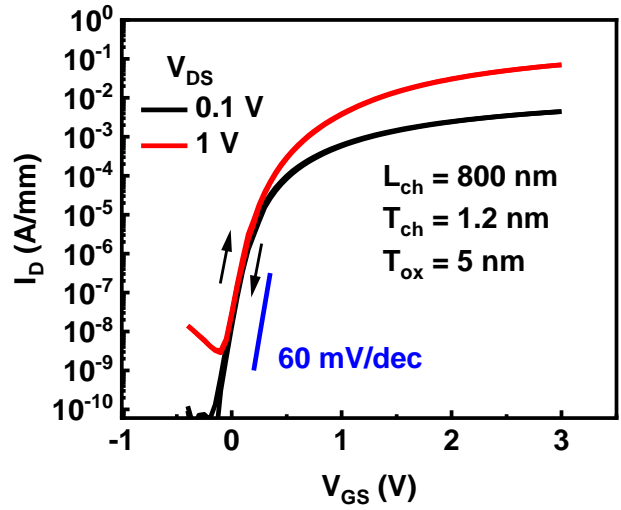

**Extended Data Fig. 5 | Performance of ALD In$_2$O$_3$ transistors with steep subthreshold slope.**

I$_D$-V$_{GS}$ characteristics of an ALD In$_2$O$_3$ transistor with L$_{ch}$ of 800 nm, T$_{ch}$ of 1.2 nm and 5 nm HfO$_2$ as the gate oxide.

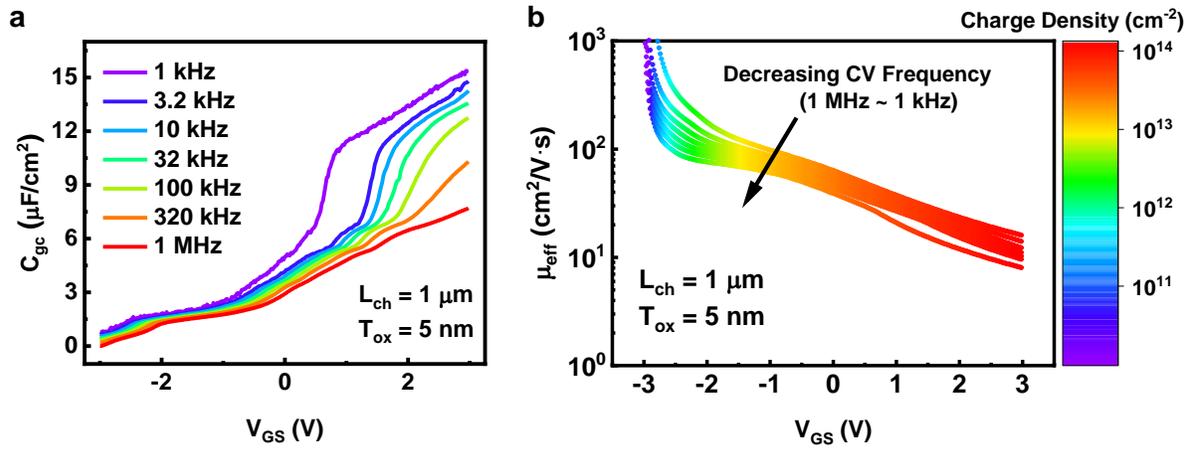

**Extended Data Fig. 6 | Split C-V measurements. a,** Measured gate to channel capacitance, normalized and tared to zero at $V_{GS} = -3$ V. **b,** Effective mobility and channel charge density as a function of gate voltage calculated from the measured output characteristics and split C-V data. The carrier density estimated in the on-state ranges from high $10^{13}$ /cm$^2$ up to above $10^{14}$ /cm$^2$ in deep accumulation. The effective mobility decreases slightly as a function of gate voltage, presumably due to increased surface roughness scattering under high surface electric field.

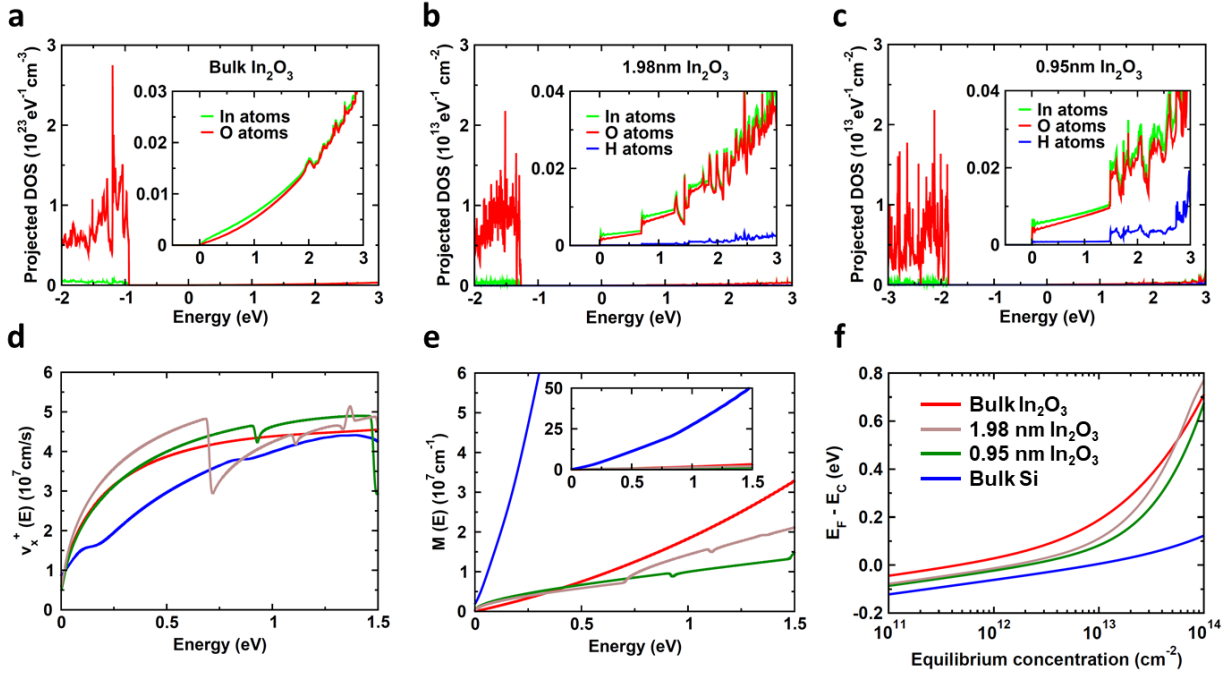

**Extended Data Fig. 7 | Electronic properties of In$_2$O$_3$ and Si from DFT. a-c,** Atom-projected DOS versus electron energy for bulk In$_2$O$_3$ (**a**), 1.98 nm-thick In$_2$O$_3$ (**b**), and 0.95 nm-thick In$_2$O$_3$ (**c**). **d,** Average electron velocity along the transport direction ($x$-direction) versus energy. **e,** Distribution of modes versus energy. **f,** Fermi level ($E_F$) relative to conduction band edge ($E_C$) versus equilibrium electron concentration. For panels **e-f**, the properties of bulk In$_2$O$_3$ and bulk Si are obtained assuming a thickness of 3.5 nm. All energies are relative to the conduction band edge.

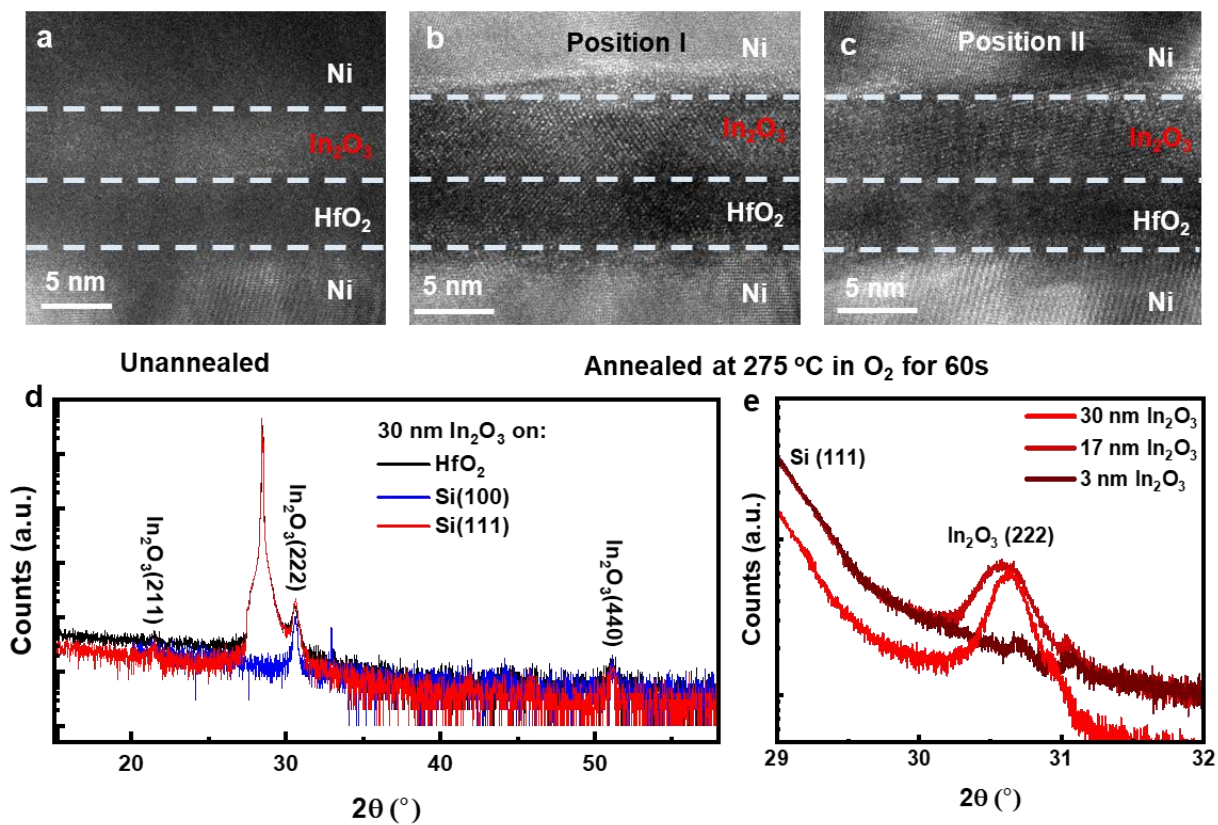

**Extended Data Fig. 8 | Crystal structure of ALD In₂O₃ thin films.** TEM cross-sectional image of ALD In$_2$O$_3$ in a Ni/In$_2$O$_3$/HfO$_2$/Ni structure at different annealing conditions, **a,** unannealed, **b,** and **c,** annealed at 275 °C in O$_2$ for 60 s at different locations. Unannealed ALD In$_2$O$_3$ has an amorphous structure while annealed ALD In$_2$O$_3$ has a polycrystalline structure. **d,** Powder XRD spectrum of thick ALD In$_2$O$_3$ with resolved peaks highlighted. Negligible differences in crystallinity are noted across different substrates. Unlabeled peaks are attributed to the substrates. The characteristic In$_2$O$_3$ (222) peak is observed by powder XRD experiment near 2θ = 30.5°. **e,** As the thickness of the In$_2$O$_3$ layer is reduced, the peaks become less sharp and less defined due in part to reduced crystallinity and in part due to instrument limitations below several nanometers. The (222) peak can be resolved down to roughly 3 nm thick layers, indicating at least some degree of partial crystallization. A small satellite peak can be observed near 2θ = 31°; its symmetric counterpart may be obscured by the nearby Si (111) peak.

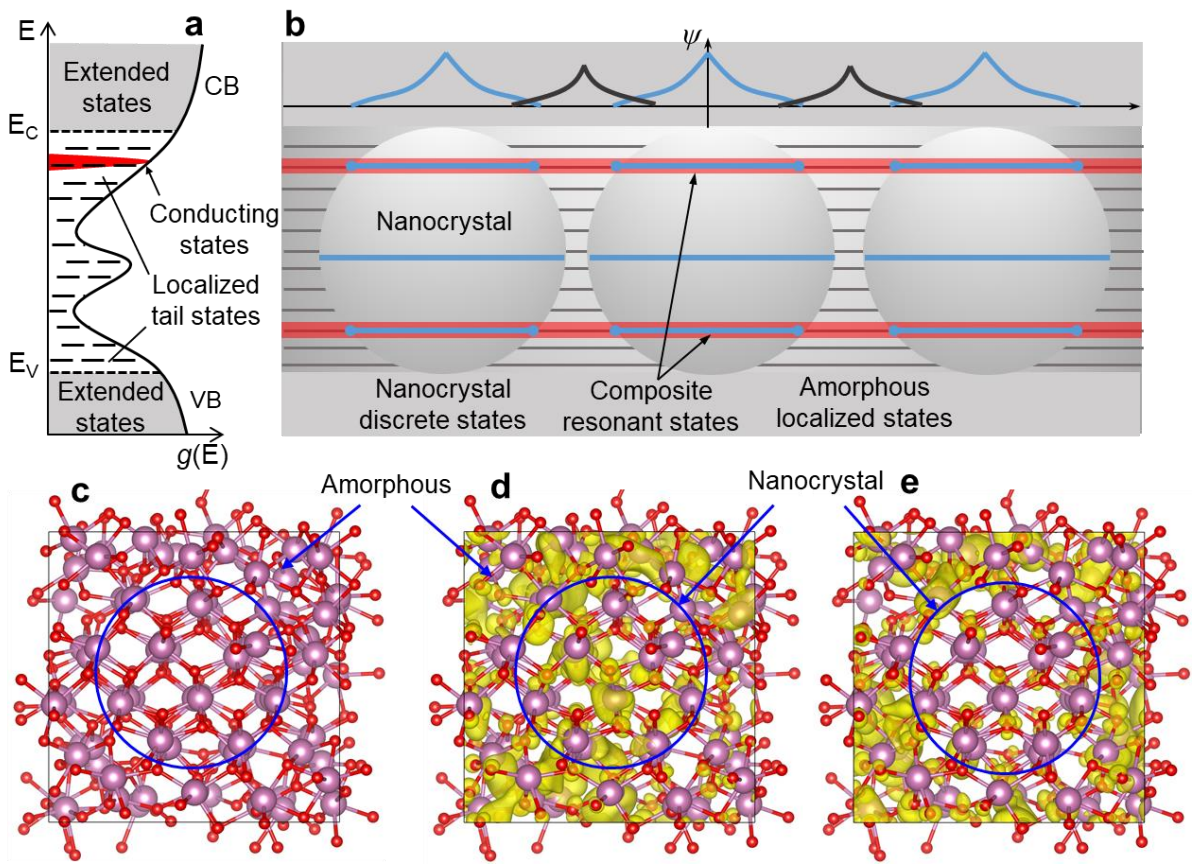

**Extended Data Fig. 9 | Charge transport modelling in ALD In$_2$O$_3$ polycrystalline films.** A In$_2$O$_3$ polycrystal film was modeled by In$_2$O$_3$ nanocrystals embedded in an In$_2$O$_3$ amorphous matrix, since the polycrystalline grain boundaries are not atomically sharp but rather amorphous level rough. **a,** Schematic representation of the density of states $g(E)$ versus energy E for an amorphous semiconductor. Localized tail states and extended states are indicated. **b,** Schematic illustration of the formation of extended states in the composite phase due to the energy-matching resonance between amorphous tail states and nanocrystal discrete quantized states. The resonant energy level is colored red, which is also indicated in (a). The wavefunction of the resonant conducting state is also shown to illustrate how wavefunction overlap leads to the delocalized conducting states. The energy-matching resonant states form the electronic transport channels throughout the In$_2$O$_3$ polycrystalline film, which results in enhanced carrier transport and electron

mobility. **c,** The atomic structure of the In$_2$O$_3$ nanocomposite phase with the In$_2$O$_3$ nanocrystal core in the center and amorphous domain in the periphery. **d** and **e,** The partial charge distribution ($|\psi|^2$) of the composite delocalized conducting state (d) and the amorphous localized tail state (e).

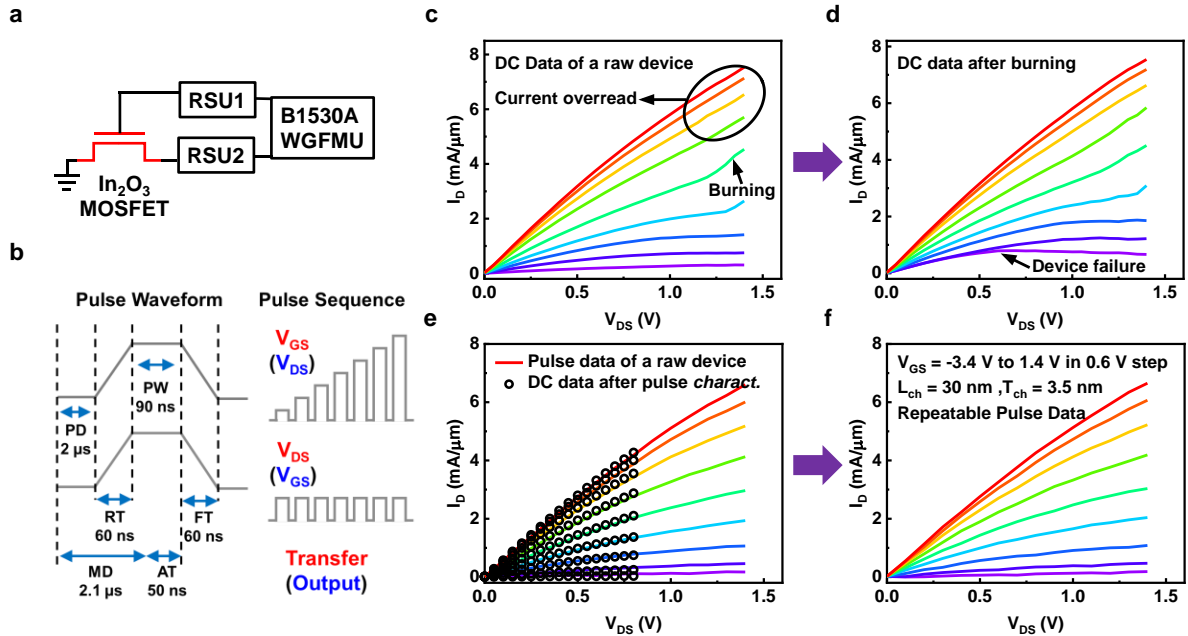

**Extended Data Fig. 10 | Pulsed I-V measurement of ALD In$_2$O$_3$ transistors. a,** Schematic diagram of the pulsed I-V measurement setup. **b,** Pulse waveform of transfer and output measurements. **c,** DC measurement of output curve at high drain bias, showing strong current drifting due to device burning. **d,** A second DC measurement of the same output curve at high drain bias, showing unrecoverable device degradation. **e,** DC measurement at low drain bias and pulsed I-V measurement at high drain bias of the same output curve, without device degradation. **f,** A second pulsed I-V measurement of the same output curve at high drain bias, showing no device degradation. c-f are data of ALD In$_2$O$_3$ transistors with $L_{ch}$ of 30 nm, $T_{ch}$ of 3.5 nm and 3 nm HfO$_2$ as the gate insulator.

**Extended Data Table 1 | Performance of representative devices with different semiconductor materials**

| Material | $L_{ch}$ (nm) | $I_{D,max}$ (A/mm) | $V_{DS}$ (V) | $G_m$ (S/mm) | $V_{DS}$ (V) | Reference |
|---|---|---|---|---|---|---|
| GaN | 50.00 | 3.60 | 10.00 | 0.60 | 5.00 | [21] A. HICKMAN et al., JEDS, 2021 |
| GaN | 20.00 | 3.00 | 4.00 | 1.36 | 4.00 | [22] Y. Tang et al., EDL, 2015 |
| GaN | 20.00 | 4.00 | 4.00 | 1.50 | 4.00 | [2] K. Shinohara et al., IEDM, 2012 |
| GaN | 60.00 | 3.30 | 4.00 | 1.50 | 4.00 | [2] K. Shinohara et al., IEDM, 2012 |
| GaN | 40.00 | - | - | 2.20 | - | [2] K. Shinohara et al., IEDM, 2012 |
| GaN | 80.00 | 2.80 | 12.00 | - | - | [23] M. Qi et al., APL, 2017 |
| InGaAs | 50.00 | 3.00 | 1.00 | 1.17 | 0.50 | [24] Y. Yonai et al., IEDM 2011 |
| InAs | 87.00 | 1.20 | 0.90 | 3.00 | 0.90 | [25] H. B. Jo et al., EDL, 2018 |
| InGaAs | 70.00 | 1.30 | 0.50 | 3.45 | 0.50 | [26] J. Lin et al., EDL, 2016 |
| InGaAs | 130.00 | 2.03 | 0.60 | - | - | [27] X. Zhou et al., APE, 2012 |
| Graphene | 2500.00 | 5.00 | 3.00 | 2.00 | 2.20 | [3] Y. Wu et al., IEDM, 2011 |
| Graphene | 2000.00 | 3.00 | 5.00 | 0.20 | 5.00 | [28] J. Moon et al., EDL, 2009 |
| Graphene | 200.00 | 3.32 | 1.00 | 1.27 | 1.00 | [18] L. Liao et al., Nature, 2010 |
| Graphene | 500.00 | 1.00 | 1.20 | 1.20 | 1.20 | [19] S.-J. Han et al., Nano Lett., 2011 |
| ITO | 10.00 | 1.86 | 1.00 | 1.05 | 1.00 | [20] S. Li et al., IEDM 2020 |
| ITO | 200.00 | 1.15 | 1.00 | - | - | [8] S. Li et al., Nat. Mater., 2019 |
| IGZO | 100.00 | 1.30 | 1.00 | 0.61 | 2.00 | [11] K. Han et al., VLSI, 2021 |